\documentclass{emulateapj}
\begin{document}

\title{Probing the Flare Atmospheres of M Dwarfs Using Infrared Emission Lines\footnotemark[*]}\footnotetext[*]{Based on observations obtained with the Apache Point Observatory 3.5 m
telescope, which is owned and operated by the Astrophysical Research
Consortium.}

\author{Sarah J. Schmidt\altaffilmark{1,2}, Adam F. Kowalski\altaffilmark{1,2}, Suzanne L. Hawley\altaffilmark{1,2}, Eric J. Hilton\altaffilmark{1,2,3}, John P. Wisniewski\altaffilmark{1,2}, Benjamin M. Tofflemire\altaffilmark{1,4}}

\altaffiltext{1} {Department of Astronomy, University of
Washington, Box 351580, Seattle, WA 98195; sjschmidt@astro.washington.edu}
\altaffiltext{2} {Guest investigator, Dominion Astrophysical Observatory, Herzberg Institute of Astrophysics, National Research Council of Canada.}
\altaffiltext{3} {Current Address: Department of Geology and Geophysics and Institute for Astronomy, University of Hawaii at Manao, Honolulu, HI 96822}
\altaffiltext{4} {Current Address: Astronomy Department, University of Wisconsin-Madison, 475 N Charter St., Madison, WI 53706}

\begin{abstract}
We present the results of a campaign to monitor active M dwarfs using infrared spectroscopy, supplemented with optical photometry and spectroscopy.  We detected 16 flares during nearly 50 hours of observations on EV Lac, AD Leo, YZ CMi, and VB8. The three most energetic flares also showed infrared emission, including the first reported detections of P$\beta$, P$\gamma$, He  I $\lambda$10830\AA~and Br$\gamma$ during an M dwarf flare. The strongest flare ($\Delta u$ = 4.02 on EV Lac) showed emission from H$\gamma$, H$\delta$, He I $\lambda$4471\AA, and Ca II K in the UV/blue and P$\beta$, P$\gamma$, P$\delta$, Br$\gamma$, and He I $\lambda$10830\AA~in the infrared. The weaker flares ($\Delta u$ = 1.68 on EV Lac and $\Delta U$ = 1.38 on YZ CMi) were only observed with photometry and infrared spectroscopy; both showed emission from P$\beta$, P$\gamma$, and He I $\lambda$10830\AA. The strongest infrared emission line, P$\beta$, occurred in the active mid-M dwarfs with a duty cycle of $\sim$3-4\%. To examine the most energetic flare, we used the static NLTE radiative transfer code RH to produce model spectra based on a suite of one-dimensional model atmospheres. Using a hotter chromosphere than previous one-dimensional atmospheric models, we obtain line ratios that match most of the observed emission lines. 
\end{abstract}
 

\section{Introduction}
\label{sec:intro}
M dwarfs are notorious for dramatic flares, presumably caused by magnetic reconnection in their atmospheres. Weaker analogs of these flares are present on the Sun, where the surface magnetic field is thought to be powered by the rotationally induced shear between the radiative and convective layers of the Sun \citep{Parker1955}. For fully convective M dwarfs \citep[spectral types M3 and later;][]{Chabrier1997}, strong magnetic fields are created and sustained solely through turbulence and rotation within the star \citep[e.g.,][]{Browning2008}. The dynamo powering the magnetic fields which result in flares on these M dwarfs is not fully understood, but additional observations constraining the chromospheric heating are essential to a coherent picture relating stellar magnetic fields to flares. 

The atmospheric heating during a flare results in emission from many wavelength regimes, and flares have been well-observed in the X-ray \citep[e.g.,][]{Osten2010}, ultraviolet \citep[e.g.,][]{Robinson2005,Hawley2007}, optical \citep[e.g.,][]{Kowalski2009,Walkowicz2011} and radio \citep[e.g.,][]{Stepanov2001,Osten2008}. The combination of observations at these different wavelengths, especially when obtained as part of multi-wavelength flare monitoring campaigns, has informed our interpretation of the physics underlying these dramatic emission events \citep[e.g.,][]{Hawley2003,Osten2005}. However, to date there have been no concerted efforts to observe infrared emission lines from flaring stars. Quiescent M dwarfs are particularly bright in the near-infrared portion of the spectrum, so emission from the hydrogen Paschen and Brackett series and the He I $\lambda$10830\AA~transition are both easily observable and essential to probing different atmospheric heights. These infrared emission lines lines are particularly useful for examining accretion in T Tauri stars \citep[e.g.][]{Bary2008,Vacca2011}. 

In quiescent (not flaring) active M stars, high resolution (R $>$ 20,000) spectra have shown weak absorption from P$\beta$ in AU Mic \citep{Short1998a};  P$\epsilon$ was also seen in absorption in six out of ten active M dwarfs \citep{Houdebine2009}. Emission from higher-order Paschen lines has only been detected in a few serendipitous observations at the far red end of optical spectra. The first occurred during a survey to classify photometrically selected late-M and L dwarfs. \citet{Liebert1999} observed Paschen emission (P$\delta$ - P11) between 8800\AA~and 10500\AA~in a R$\sim$4300 spectrum of the M9.5 dwarf 2MASSW J0149090+295613. The flare also showed a variety of optical emission lines, but had no evidence of continuum emission. \citet{Schmidt2007} subsequently observed Paschen emission lines in a R$\sim$2000 spectrum of the M7 dwarf 2MASS J1028404-143843. The flaring spectrum included strong continuum enhancement of the entire spectrum blueward of 9200\AA~in addition to many emission lines. P$\delta$ - P11 were again identified, with equivalent widths (EW) of 2-5\AA. For both of these observations, there was no corresponding photometry, which prohibited the characterization of the overall strength and duration of the flare. 

\citet{Fuhrmeister2008} observed P$\delta$ - P11 on the M5.5 dwarf CN Leo during a large-amplitude flare with a total duration of about 45 minutes. Their data included R$\sim$40,000 spectra over the range 3000\AA~to 10500\AA. Line strengths were not given for the Paschen lines, but inspection of the 3 consecutive 100 s exposures near the peak of the flare shows a decay in line strength. \citet{Fuhrmeister2010} used one-dimensional atmosphere models to examine the emission from the flare, finding that a single model can reproduce most, but not all, flare emission lines. 

An unexplored region of the spectrum during M dwarf flares, both in observations and modeling, is the 1.0 to 2.5$\mu$m range, which contains the lower order Paschen lines, higher-order Brackett and Pfund lines, and He I $\lambda$10830\AA. We report on the results of our campaign to observe active M dwarfs in this wavelength regime with simultaneous photometric monitoring. The data include 3 flares with infrared line emission; observations of the strongest flare also include blue optical spectra. Using these data, we quantify the duty cycle for infrared flare line emission and examine the relative line strengths during the evolution of each of the three flares. Congruent with previous studies \citep[e.g.,][]{Hawley1992,Walkowicz2008a,Fuhrmeister2010}, we use one-dimensional atmospheres and  the static NLTE radiative transfer code RH \citep{Uitenbroek2001} to model the line flux ratios of the largest flare.

In Section~\ref{sec:obs} we discuss our targets and observations and in Section~\ref{sec:flareID} we describe our methods for flare identification. Individual flares are examined in Section~\ref{sec:flareall} together with our duty cycle estimate. Empirical atmospheric models that produce infrared line emission are discussed in Section~\ref{sec:mod}.

\section{Observations}
\label{sec:obs}
Our targets include three well-known mid-M flare stars; AD Leo, EV Lac, and YZ CMi, in addition to one active late-M dwarf, VB 8. Magnitudes, coordinates, and the duration of our observations for each target are given in Table~\ref{tab:targ}. To both detect IR emission and characterize a flare, we need at least infrared spectra and one band of optical photometry, but some nights include additional data. The targets, times observed, and instruments used each night are detailed in Table~\ref{tab:obs}. 

\begin{deluxetable*}{llllllllll} \tablewidth{0pt} \tabletypesize{\scriptsize}
\tablecaption{Flare Star Observations \label{tab:targ}}
\tablehead{ \colhead{Name} &  \colhead{ST} & \colhead{J} & \colhead{K} & \colhead{t$_{\rm obs}$ (h:m) } & \colhead{N$_{\rm flares}$} &\colhead{t$_{\rm flare}$ (h:m) }  & \colhead{N$_{\rm flares}$ IR} &\colhead{t$_{\rm flare}$ IR (h:m)} & \colhead{Frac$_{\rm IR}$}}
\startdata
YZ CMi &  M4.5    & 6.58$\pm$0.02 & 5.70$\pm$0.02& 16:36   & 8 & 1:55 & 1 & 0:08 & 0.008\\
AD Leo &  M3         & 5.45$\pm$0.02 & 4.59$\pm$0.02&  12:18  & 1 & 0:31 & 0 & 0:00       & 0 \\
EV Lac   &  M3.5 & 6.11$\pm$0.03  & 5.30$\pm$0.02 & 15:21 & 6 & 3:56 & 2 & 1:14 & 0.081 \\
VB 8       &  M7      & 9.78$\pm$0.03 &  8.82$\pm$0.02 & 4:37   & 1 & 0:15 & 0 & 0:00       & 0 \\
\hline
M3-M4.5 &   &  &   & 44:15 & 15 & 6:22 & 3 & 1.37 & 0.031 \\
total &  &   &   & 48:52 & 16 & 6:38 & 3 & 1.37 & 0.028 \\
\enddata
\tablecomments{The total time observed is given in column 5 (t$_{\rm obs}$), the total time each object spent in flare is given in column 7 (t$_{\rm flare}$), and the time with observed infrared line emission is given in column 9 (t$_{\rm obs}$; see Figures~\ref{fig:evoct}, \ref{fig:evnov}, and \ref{fig:yzfeb}). The last column gives the fraction of time each object spent with infrared line emission.}
\end{deluxetable*}

\begin{deluxetable*}{lllllllll} \tablewidth{0pt} \tabletypesize{\scriptsize}
\tablecaption{List of Observations \label{tab:obs}}
\tablehead{ \colhead{UT date} & \colhead{Target} & \colhead{Telescope/Instr.} & \colhead{Filter\tablenotemark{1}} & \colhead{ET (s)} & \colhead{UT Time (h:m)} & \colhead{Time (h:m)} & \colhead{t$_{\rm obs}$ (h:m)} & \colhead{N$_{\rm flares}$}}
\startdata													
2009 Feb 4	&	YZ CMi	&	APO TripleSpec\tablenotemark{2}	&		&	4	&	3:20 to 7:10	&	3:50	&	1:42	& 1\\
2009 Feb 4	&	YZ CMi	&	ARCSAT flare-cam	&	\textbf{u}, g	&	7, 2	&	5:30 to 7:12	&	1:42	&		\\
\hline															
2009 Feb 4	&	AD Leo	&	APO TripleSpec	&		&	2	&	7:14 to 13:46	&	6:32	&	5:20& 1	\\
2009 Feb 4	&	AD Leo	&	ARCSAT flare-cam	&	\textbf{u}, g	&	5, 1	&	7:18 to 12:38	&	5:20	&		\\
\hline	
2009 Apr 14	&	AD Leo	&	APO TripleSpec	&		&	2-5	&	2:54 to 7:44	&	4:50	&	1:18	 & 0\\
2009 Apr 14	&	AD Leo	&	ARCSAT flare-cam	&	\textbf{g}	&	10-100	&	2:28 to 4:12	&	1:44	&		\\
\hline															
2009 Oct 24	&	EV Lac	&	APO TripleSpec	&		&	8	&	1:41 to 6:51	&	5:10	&	3:35& 1	\\
2009 Oct 24	&	EV Lac	&	ARCSAT flare-cam	&	\textbf{u}, g	&	10, 1	&	1:06 to 5:05	&	3:59	&		\\
2009 Oct 24	&	EV Lac	&	DAO Spectrograph\tablenotemark{3}	&		&	60-300	&	2:26 to 9:05	&	6:39	&		\\
2009 Oct 24	&	EV Lac	&	NMSU 1-m camera	&	U	&	4	&	1:26 to 9:30	&	8:04	&		\\
\hline															
2009 Oct 27	&	EV Lac	&	APO TripleSpec	&		&	8	&	1:04 to 6:53	&	5:49	&	5:19 & 3	\\
2009 Oct 27	&	EV Lac	&	ARCSAT flare-cam	&	\textbf{u}, g	&	5, 1	&	1:34 to 9:13	&	7:49	&		\\
2009 Oct 27	&	EV Lac	&	DAO Spectrograph	&		&	200-420	&	2:02 to 6:18	&	4:16	&		\\
2009 Oct 27	&	EV Lac	&	NMSU 1-m camera	&	U	&	4	&	1:12 to 9:18	&	8:06	&		\\
\hline															
2010 Apr 25	&	VB 8 	&	APO TripleSpec	&		&	60	&	7:24 to 11:07	&	3:43	&	3:12 & 0	\\
2010 Apr 25	&	VB 8 	&	ARCSAT flare-cam	&	\textbf{g}	&	120	&	7:57 to 11:06	&	2:33	&		\\
\hline														
2010 May 26	&	VB 8 	&	APO TripleSpec	&		&	30	&	7:28 to 8:53	&	1:25	&	1:25	& 1\\
2010 May 26	&	VB 8 	&	ARCSAT flare-cam	&	\textbf{g}	&	60	&	6:22 to 8:54	&	2:32	&		\\
\hline															
2010 May 26	&	EV Lac	&	APO TripleSpec	&		&	8	&	9:01 to 11:39	&	2:38	&	1:50	& 1\\
2010 May 26	&	EV Lac	&	ARCSAT flare-cam	&	\textbf{u}, g	&	10, 2	&	9:10 to 11:00	&	1:50	&		\\
\hline	
2010 Nov 27	&	EV Lac	&	APO TripleSpec	&		&	5	&	0:24 to 5:14	&	4:50	&	4:37& 1	\\
2010 Nov 27	&	EV Lac	&	ARCSAT flare-cam	&	\textbf{u}, g, r	&	5, 1, 1	&	0:37 to 5:17	&	4:40	&	\\
\hline															
2011 Feb 14	&	YZ CMi	&	APO TripleSpec	&		&	4	&	2:07 to 6:08	&	4:01	&	4:01	& 4\\
2011 Feb 14	&	YZ CMi	&	ARCSAT flare-cam	&	g, r, i	&	1, 1, 1	&	1:24 to 4:30	&	3:06	&		\\
2011 Feb 14	&	YZ CMi	&	NMSU 1-m camera	&	\textbf{U}	&	10	&	2:07 to 7:20	&	5:13	&		\\
\hline															
2011 Feb 15	&	AD Leo	&	APO TripleSpec	&		&	2	&	9:35 to 12:28	&	2:53	&	2:36 & 0	\\
2011 Feb 15	&	AD Leo	&	ARCSAT flare-cam	&	g, r, i	&	1, 1, 1	&	9:26 to 10:36	&	1:10	&		\\
2011 Feb 15	&	AD Leo	&	NMSU 1-m camera	&	\textbf{U}	&	4	&	9:52 to 13:09	&	3:17	&		\\
\hline															
2011 Feb 15	&	YZ CMi	&	APO TripleSpec	&		&	4	&	1:16 to 3:08	&	1:52	&	1:07 & 2	\\
2011 Feb 15	&	YZ CMi	&	ARCSAT flare-cam	&	g, r, i	&	1, 1, 1	&	1:26 to 3:56	&	2:30	&		\\
2011 Feb 15	&	YZ CMi	&	APO TripleSpec	&		&	4	&	4:49 to 9:34	&	4:45	&	4:45	\\
2011 Feb 15	&	YZ CMi	&	ARCSAT flare-cam	&	g, r, i	&	1, 1, 1	&	4:50 to 5:27	&	1:37	&		\\
2011 Feb 15	&	YZ CMi	&	NMSU 1-m camera	&	\textbf{U}	&	10	&	1:59 to 9:41	&	7:42	&		\\
\hline															
2011 Feb 21	&	AD Leo	&	APO TripleSpec	&		&	2	&	8:41 to 12:16	&	3:35	&	3:04& 0	\\
2011 Feb 21	&	AD Leo	&	ARCSAT flare-cam	&	\textbf{u}, g, r	&	5, 1, 1	&	6:07 to 11:45	&	5:38	&	\\
\hline															
2011 Feb 22	&	YZ CMi	&	APO TripleSpec	&		&	4	&	2:08 to 7:09	&	5:01	&	5:01 &1	\\
2011 Feb 22	&	YZ CMi	&	ARCSAT flare-cam	&	\textbf{u}, g, r, i	&	6, 1, 1, 1	&	1:49 to 7:40	&	5:51	&	\\
\enddata
\tablecomments{Column 7 (Time) refers to the total duration of observations for each instrument, while column 8 (T$_{\rm obs}$) gives the duration of observations that overlap for the photometry and infrared spectroscopy. Column 9 (N$_{\rm flares}$) gives the number of flares found from the photometry using the process described in Section~\ref{sec:flareID}.}
\tablenotetext{1}{The filter in bold was used to identify flares for those observations.}
\tablenotetext{2}{R$\sim$3500; $\lambda \sim 0.95$ to $2.45 \mu$m; see section~\ref{sec:obst}}
\tablenotetext{3}{R$\sim$750; $\lambda \sim$ 3540 to 4710\AA; see section~\ref{sec:obsd}}
\end{deluxetable*}

\subsection{Flare-cam on the ARCSAT 0.5-m}
\label{sec:obsf}
Photometry for all of the nights was obtained using Flare-cam on the Astrophysical Research Consortium Small Aperture Telescope (ARCSAT). ARCSAT was formerly used as the photometric calibrating telescope for the Sloan Digital Sky Survey \citep{York2000,Tucker2006}. Its location at Apache Point Observatory (APO) makes it an ideal telescope for obtaining simultaneous data with the ARC 3.5-m telescope, which we used for our infrared spectroscopy (see Section~\ref{sec:obst}). Flare-cam is equipped with $ugri$ filters, and the CCD is optimized for observing flare stars because of its good blue response and fast readout \citep{Hilton2011phd}. Exposure times and filters are given in Table~\ref{tab:obs}. 

The data were reduced using standard IRAF routines combined with a custom python code that tracked the change in each star's position over the course of the night. The magnitudes were calibrated using differential photometry with respect to the brightest stars in the image.  See \citet{Hilton2011phd} for more details on the photometric reductions.

\subsection{NMSU 1-m}
\label{sec:obsn}
For four of the eleven nights of observations listed in Table~\ref{tab:obs}, we also obtained $U$-band photometry using the NMSU 1-m, a robotically operated telescope located at APO \citep{Holtzman2010}. Typical exposure times were 4 -10 seconds, and readout was 10 seconds. The reductions were performed using an automated pipeline which measured the magnitude of the flare star with respect to several background stars. 

\subsection{TripleSpec on the ARC 3.5-m}
\label{sec:obst}
Infrared spectra were obtained with the TripleSpec instrument on the ARC 3.5-m telescope at APO. TripleSpec is a cross-dispersed near infrared spectrograph that covers 0.95 to 2.45 microns \citep{Wilson2004}. We used the 1.1\arcsec~slit, resulting in a resolution of R$\sim$3500. All data were obtained using an A/B nod pattern, shifting the star along the slit in order to perform sky subtraction. We obtained data for an A0 calibrator star every $\sim$40 minutes in order to correct for the changing telluric absorption over the course of the night. Our typical exposure times were 2-4 seconds for AD Leo, EV Lac, and YZ CMi, and 30 seconds for VB 8.

The data were reduced using a version of SpexTool modified to work with ARC 3.5-m TripleSpec data \citep{Cushing2004}. We constructed telluric correction spectra from our A0 standards using the routine included in SpexTool \citep{Vacca2003}, but modified the remaining post-processing routines to automatically process each spectrum instead of using the provided GUI interface. Although the formal residuals of our wavelength solution were 0.5-1 pixel, the curved, tilted orders of the spectra impose additional systematic effects in the wavelength calibration. 

\begin{figure}
\includegraphics[width=\linewidth]{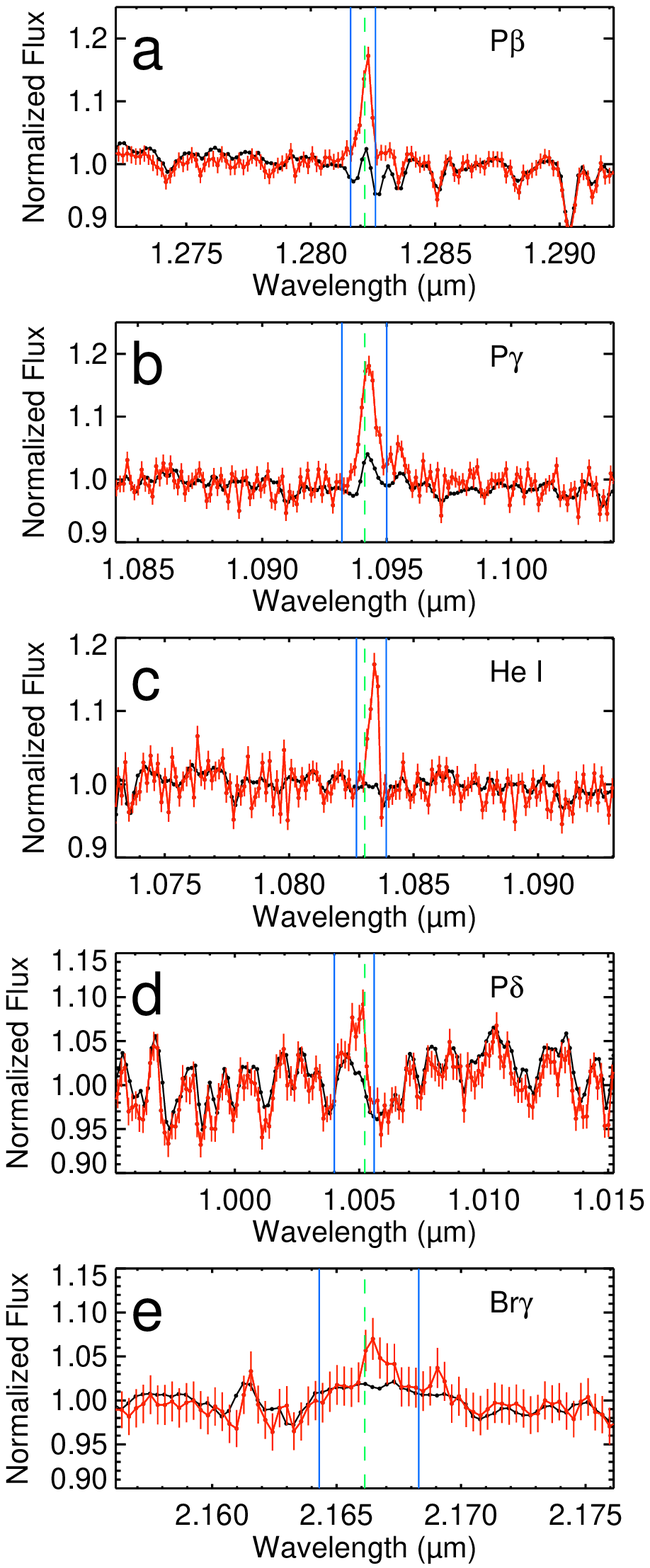} 
\caption{Spectra of EV Lac during quiescence (black) co-added from $\sim$2 hours of exposure during UT 2009 October 24, and from 8 co-added spectra with exposure times of 8 seconds each during the flare peak on UT 2009 October 27 (red). The emission lines are labelled and their central wavelengths are shown (dashed green lines). The regions used to measure the EW of each line are also shown (solid blue lines). The shifts of some lines from the nominal central wavelengths are likely due to small (1-2 pixel) systematic errors in the wavelength solution (see text).} \label{fig:spec}
\end{figure}

We detected P$\beta$, P$\delta$, P$\gamma$, Br$\gamma$, and He I $\lambda$10830\AA~during the most energetic flare observed. No higher order Brackett or Pfund emission was detected in any of our spectra. We measured the equivalent widths (EW) using regions defined individually for each line in order to include all observed flux; for P$\beta$, P$\delta$, P$\gamma$, and He I $\lambda$10830\AA~these were 10-20\AA~wide (6-12 pixels; 0.001-0.002 $\mu$m) and for Br$\gamma$ the line region was 40\AA~wide (14 pixels; 0.004 $\mu$m). Continuum regions were defined as $\pm$0.01$\mu$m on either side of each line. Quiet and flare profiles for the five lines are shown with the regions used for line measurements in Figure~\ref{fig:spec}. The spectrum surrounding and underlying each of the emission lines contains many other molecular and atomic features so the EWs of the emission lines are not zero even in quiescence. The EW measured in the quiescent spectrum is subtracted from each flare measurement.

We could not determine absolute line fluxes directly from individual TripleSpec spectra because the observed flux in each spectrum varies due to the movement of the star on and off the slit during the nod pattern. We used a method similar to the $\chi$ factor of \citet{Walkowicz2004} to calculate absolute line flux by multiplying measured equivalent widths (which do not depend on the continuum level) by a calibrated continuum flux.  Continuum fluxes were obtained from a quiet, co-added, high S/N spectrum of each star normalized to 2MASS photometry using 2MASS filter curves \citep{Cohen2003,Skrutskie2006}. While this method would not be feasible in the UV and optical due to white light emission, continuum enhancement during flares is negligible in the $JHK_S$ passbands. \citet{Davenport2011} use a flare continuum model on an M3 star to predict that a flare with $\Delta u =$ 4 mags would produce a $\Delta J <$ 10 millimags peak, and \citet{Tofflemire2011} report no broad-band ($J$, $H$, $K_{S}$) continuum enhancements above a level of 5-8 millimags during flares having similar total energy as reported here. The variations detected by \citet{Tofflemire2011} and \citet{Davenport2011} are smaller than the formal uncertainty quoted with the 2MASS magnitudes (20-30 millimags), so we assume that the variation between the published 2MASS magnitudes and the magnitudes of the M dwarfs during our observing was negligible. 

\subsection{DAO 1.8-m}
\label{sec:obsd}
For two nights, we used the DAO 1.8-m telescope with the SITe5 CCD and spectrograph to observe EV Lac during a coordinated campaign with the telescopes at APO. Our setup resulted in a spectral resolution of R$\sim$750 and wavelength coverage from 3550\AA-4700\AA. We measured Ca II K, He I $\lambda$4471\AA, and the Hydrogen Balmer series H$\gamma$ and H$\delta$. Exposure times for EV Lac ranged from 60 to 420 seconds. Due to these relatively long integration times, additional cosmic ray cleaning was performed with the LACOSMIC utility \citep{van-Dokkum2001}. 

The spectra were wavelength-calibrated with a FeAr lamp and flux-calibrated using data from the standard star G191B2B, then spectrophotometrically calibrated by normalizing to the simultaneous $U$-band data. Equivalent widths are not useful for blue flare spectra because of the changes in the surrounding continuum flux during the flare. Instead, we measured absolute line fluxes directly from the data. The values we use during the flare have the quiet line flux subtracted. 

\section{Identifying Flares}
\label{sec:flareID}
Flares are most easily seen at blue and ultraviolet wavelengths, where the hot, white-light continuum emission from the flare is in high contrast to the small amount of flux emitted from cool M dwarf photospheres \citep{Lacy1976,Hawley1991}. To identify as many flares as possible, we used the bluest band of photometry available. This was typically $u$, but for some nights only $U$ was available, and VB 8 was too faint to observe in $U$ or $u$, so we used $g$-band data. The band used to identify flares for each set of observations is given in Table~\ref{tab:obs}.

\begin{figure}
\includegraphics[width=\linewidth]{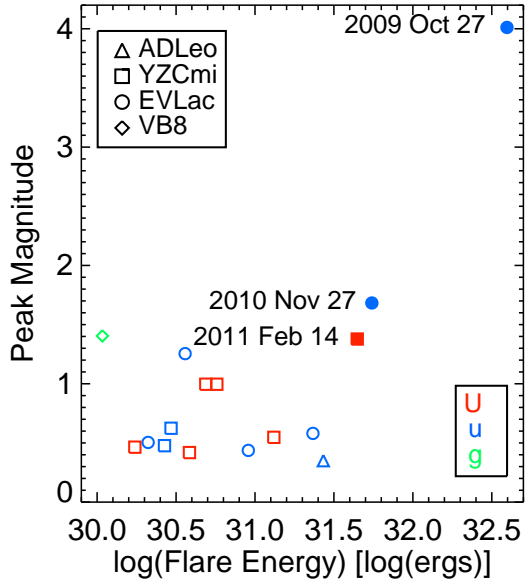} 
\caption{Peak magnitude in $U$-, $u$-, or $g$-band as a function of total flare energy in those bands. The flares from AD Leo (triangles), YZ CMi (squares), EV Lac (circles) and VB 8 (diamonds) observed in $U$ (red), $u$ (blue), and $g$ (green) band are shown. The flares where IR line emission was observed are distinguished (solid symbols) and labelled with their dates. The flares with accompanying IR line emission have the largest total flare energy and relatively high peak magnitudes.} \label{fig:chfl}
\vspace{8pt}
\end{figure}

Photometrically, flares are observed as excursions above the mean quiescent value of the star's flux, which can be any size or shape. Realistically, flare detection must take into account small variations in the continuum caused by observational effects and so a minimum duration and energy above the observed quiescent value is required. To identify individual flares, we used the custom IDL code discussed in \citet{Hilton2011phd}, which selects peaks that have at least three consecutive epochs more than 3 standard deviations above the local quiescent light curve. At least one of those epochs must be 5 sigma above quiescence. We reviewed each flare by eye to confirm that the deviations from the mean were not caused by bad photometry. Over the course of 48.9 hours of observations on four different stars, we observed a total of 16 flares, which are listed per star in Table~\ref{tab:targ} and per night in Table~\ref{tab:obs}. Figure~\ref{fig:chfl} shows the energy and peak magnitude of each flare.  

To identify flares which had associated IR line emission, we examined the measured EWs of P$\beta$ and He I $\lambda$10830\AA~as a function of time during the flare. We found that the three most energetic flares, which occurred on EV Lac on UT 2009 October 27 and UT 2010 November 27, and on YZ CMi on UT 2011 February 14, each showed infrared emission lines. These flares are discussed in detail in Section~\ref{sec:flare}.

\section{Characterizing Infrared Flares}
\label{sec:flareall}
\subsection{Individual Flares}
\label{sec:flare}
\begin{figure*}
\includegraphics[width=0.95\linewidth]{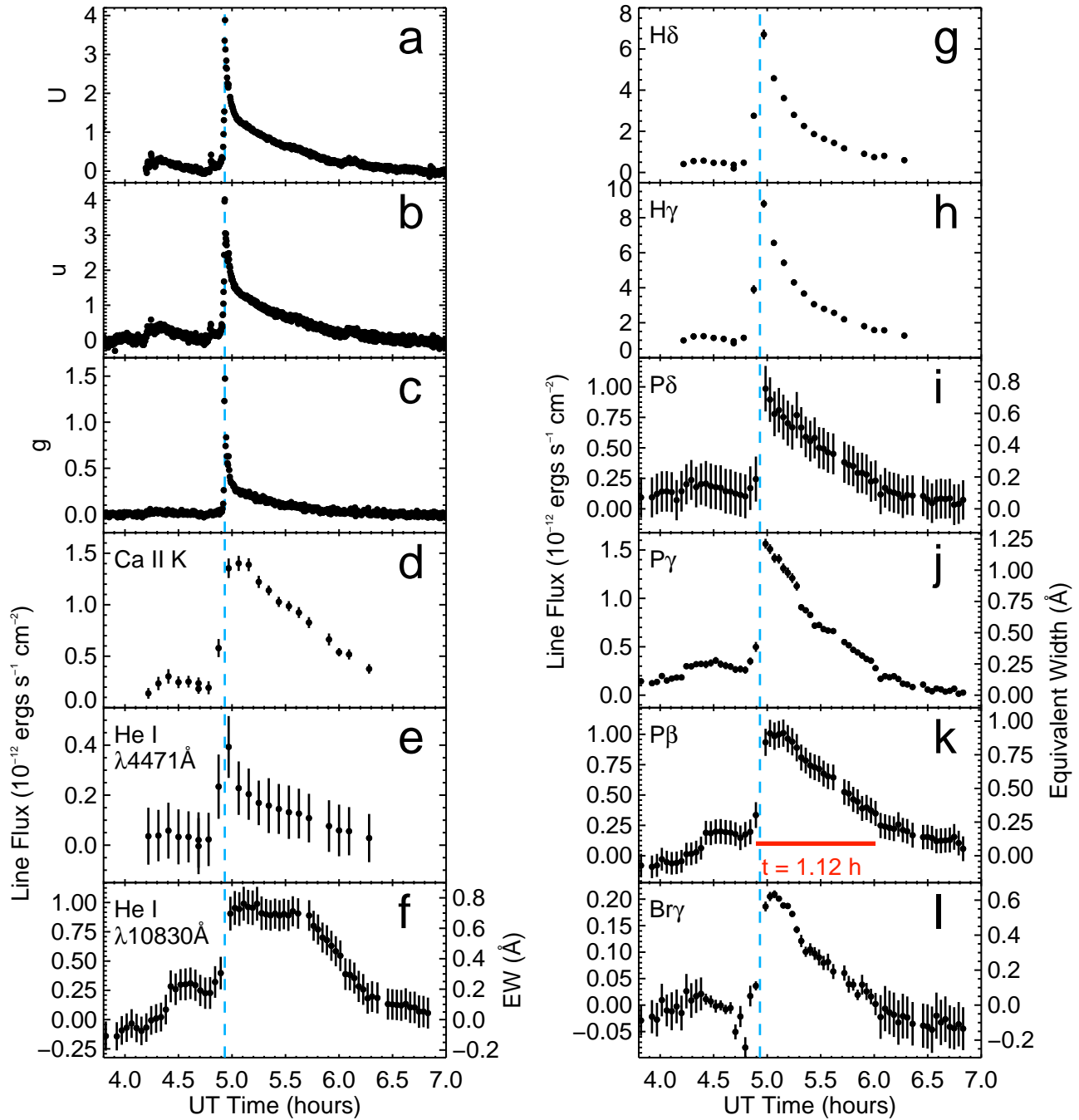} 
\caption{Data from the UT 2009 October 27 flare observed on EV Lac. Panels (a), (b), and (c) show $U$, $u$, and $g$-band light curves during the flare. The panels (d) - (l) show the variations in the optical and infrared lines. The flaring line flux (with a quiescent value subtracted) is shown along the left y axis, and for the infrared data equivalent width is shown along the right y axis. Panel (k) shows the time used as the flare duration in Table~\ref{tab:targ} (red line). The time of the peak $u$-band flux is shown in each panel (blue dashed line). The infrared emission line fluxes are averaged over 8 exposures of 8 seconds each. The total exposure time for each point is 64 seconds, but including instrumental overhead and time to execute the nod pattern, the average cadence is 2.5 minutes. Small gaps in the infrared line data are due to standard star observations.} \label{fig:evoct}
\end{figure*}

\textbf{2009 October 27 flare on EV Lac:} We were observing with all four instruments during the most energetic event, a $\Delta u$ = 4.02\footnote[2]{Although the $\Delta u$ represents a negative change in magnitude (corresponding with an increase in flux) we adopt a convention of $\Delta u = | u_{flare} - u_{quiet}|$ throughout.} magnitude flare on EV Lac on UT 27 October 2009. The light curves for our observations are shown in Figure~\ref{fig:evoct}. The photometry (in $U$-, $u$-, and $g$-band) exhibits a typical flare light curve with a fast rise and exponential decay. The $u$-band flare emission lasted 1.68 hours and released a total energy of 3.9 $\times$10$^{32}$ ergs.

\begin{figure}
\includegraphics[width=0.95\linewidth]{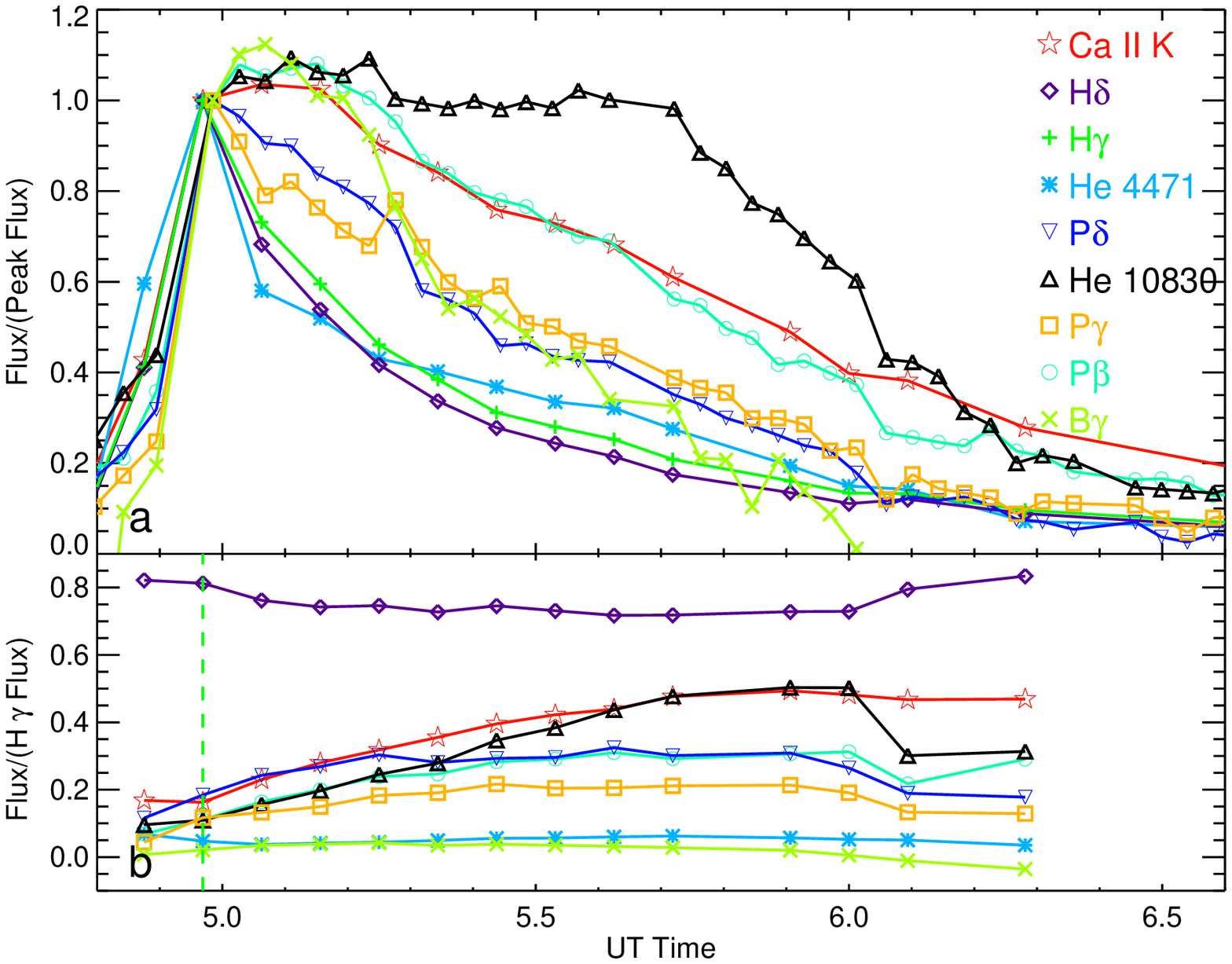} 
\caption{Line fluxes for the nine optical and infrared emission lines detected during the UT 2009 October 27 flare observed on EV Lac. In panel (a), the line fluxes are normalized to their flux at the time of peak emission for H$\gamma$ (t = 4.97 hours). The normalization emphasizes the different decay patterns. Panel (b) shows the ratios of the other eight emission lines to the H$\gamma$ flux. The IR lines were binned to the same time resolution as the optical lines. The vertical green dashed line shows the location of the flare peak in H$\gamma$.} \label{fig:linec}
\end{figure}

The combination of optical (DAO) and infrared (TripleSpec) spectroscopy allows us to examine a total of nine emission lines - H$\gamma$, H$\delta$, He I $\lambda$4471\AA, and Ca II K in the UV/blue part of the spectrum, and P$\beta$, P$\gamma$, P$\delta$, Br$\gamma$, and He I $\lambda$10830\AA~in the infrared. Figure~\ref{fig:linec} shows the lightcurve of each emission line normalized to its value at t = 4.97 hours (the peak of $u$-band emission), and the ratio of each line to H$\gamma$ for comparison of their evolution during the flare. 

The light curves for H$\gamma$, H$\delta$, and He I $\lambda$4471\AA~have a fast-rise exponential-decay shape similar to the photometry.  P$\gamma$ and P$\delta$ show a similar fast rise, but their decay is slower than the Balmer series lines. The P$\beta$ and Ca II K emission both peak after the other Paschen and Balmer series lines, and exhibit an even slower decay after their late peaks. Br$\gamma$ is similar to P$\beta$ and Ca II K in its late peak, but seems to decay faster than any other line. This may be an observational effect, as it is by far the weakest line detected. Without a stronger detection, we assume that its ratio to the Paschen lines is constant throughout the flare.  The He I $\lambda$10830\AA~emission shows a shape distinct from the rest of the lines -- it remains nearly at its peak flux for 0.8 hours, approximately half of the duration of the flare in $u$-band. 

The slow decay during the gradual phase is a well-known property of Ca II K \citep[e.g.\ ][]{Bopp1973,Hawley1991,Fuhrmeister2008}, but in this flare, He I $\lambda$10830\AA~emission traces a region that remains heated for an even longer portion of the gradual phase than Ca II K. This could be due to the Neupert effect, where the line responds to the total cumulative flare heating for which the time integral of the $U$-band (white light emission) is often used as a proxy \citep{Hawley1995,Osten2005}. Section~\ref{sec:mod} describes our efforts to model the emission lines from this flare.

\begin{figure}
\includegraphics[width=\linewidth]{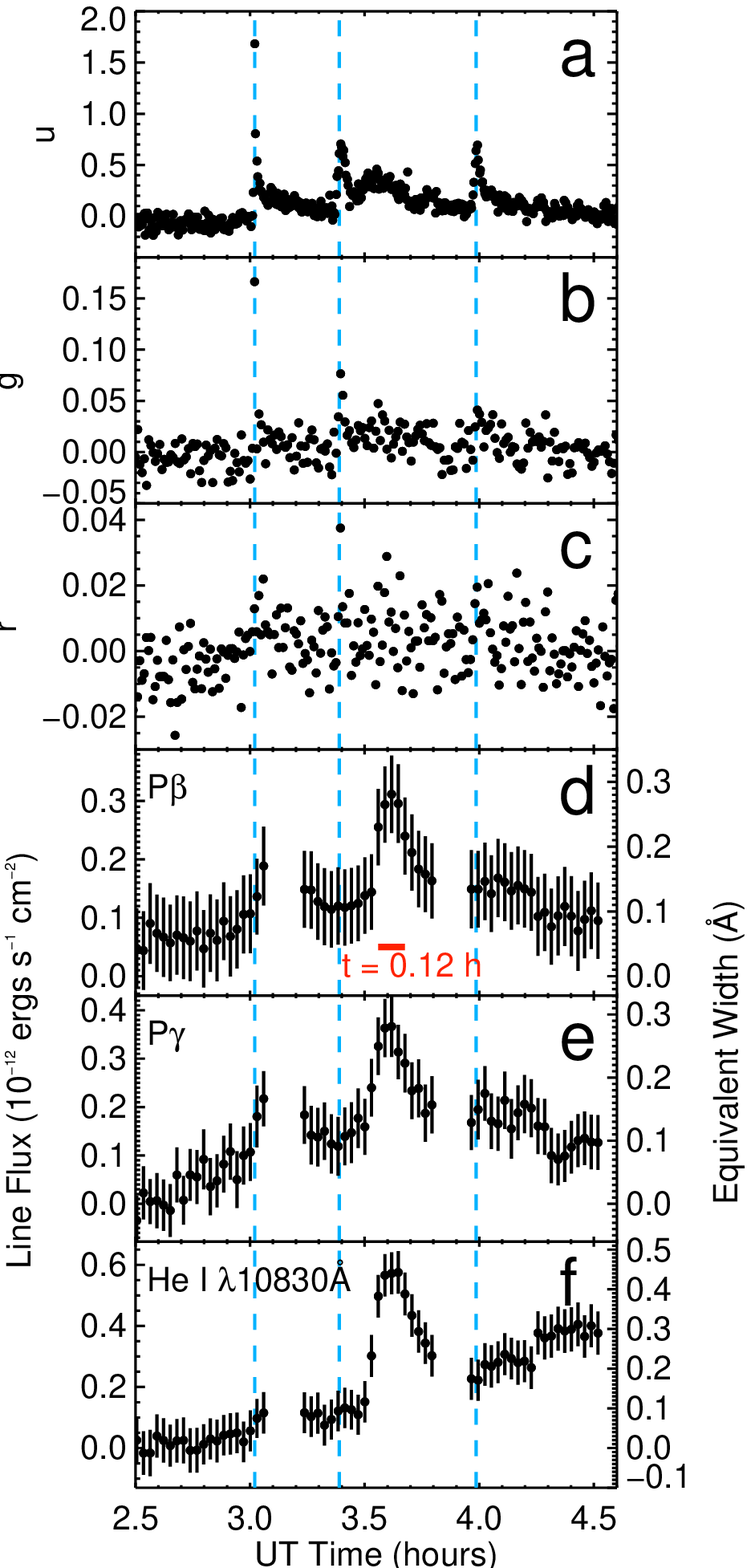} 
\caption{Data from the UT 2010 November 27 flare observed on EV Lac. Panels (a), (b), and (c) show the change in $u$-, $g$-, and $r$-bands during the flare, and panels (d), (e), and (f) show the flux (left axis) and EW (right axis) of the infrared emission lines detected during the flare. Panel (d) shows the time used as the flare duration in Table~\ref{tab:targ} (red line). The three dashed blue lines show the times when the three flare peaks occurred in the $u$-band. The infrared emission line fluxes are averaged over 8 exposures of 5 seconds each. The total exposure time for each point is 40 seconds; the average cadence is 1.7 minutes. The gaps in the infrared line data are due to standard star observations.} \label{fig:evnov}
\end{figure}

\textbf{2010 November 27 Flare on EV Lac:}
We observed another flare with infrared line emission on EV Lac on UT 2010 November 27. The flare peaked at $\Delta u$ = 1.68, and over the course of t = 1.30 hours it emitted 5.5 $\times$10$^{31}$ ergs in the $u$-band. We observed with both ARCSAT and TripleSpec during the flare, and have photometry in $g$ and $r$-band in addition to the $u$-band data. The photometry and the line flux lightcurves for P$\beta$, P$\gamma$, and He I $\lambda$10830\AA~are shown in Figure~\ref{fig:evnov}. There was no discernible emission in P$\delta$ and Br$\gamma$.

This peculiarly-shaped flare contains three separate peaks in the $u$-band photometry. After the first and third peak, the flux seems to decay exponentially, but after the middle peak there is a gentle rise in the $u$-band flux. TripleSpec was taking observations of a standard star during the first peak of the flare, so it is unknown if the emission lines showed the same fast-rise exponential-decay as the first photometric peak. The rise in P$\beta$ and P$\gamma$ line emission before and after the standard star gap suggests that those lines showed some emission between the first and second peaks of the flare. An observed increase in infrared line emission occurred $\sim$0.2 hours after the second peak in the $u$-band photometry, tracing a gentle rise and decay. 

The shape of this flare is very different than that of the UT 2009 October 27 flare on EV Lac, and the relative line strengths are also different. In the previous flare, P$\beta$, P$\gamma$ and He I $\lambda$10830\AA~emitted nearly the same peak flux. In this flare, He I $\lambda$10830\AA~peaked at twice the strength of the P$\beta$ and P$\gamma$ lines, indicating a different pattern of atmospheric heating during the two flares.

\begin{figure}
\includegraphics[width=\linewidth]{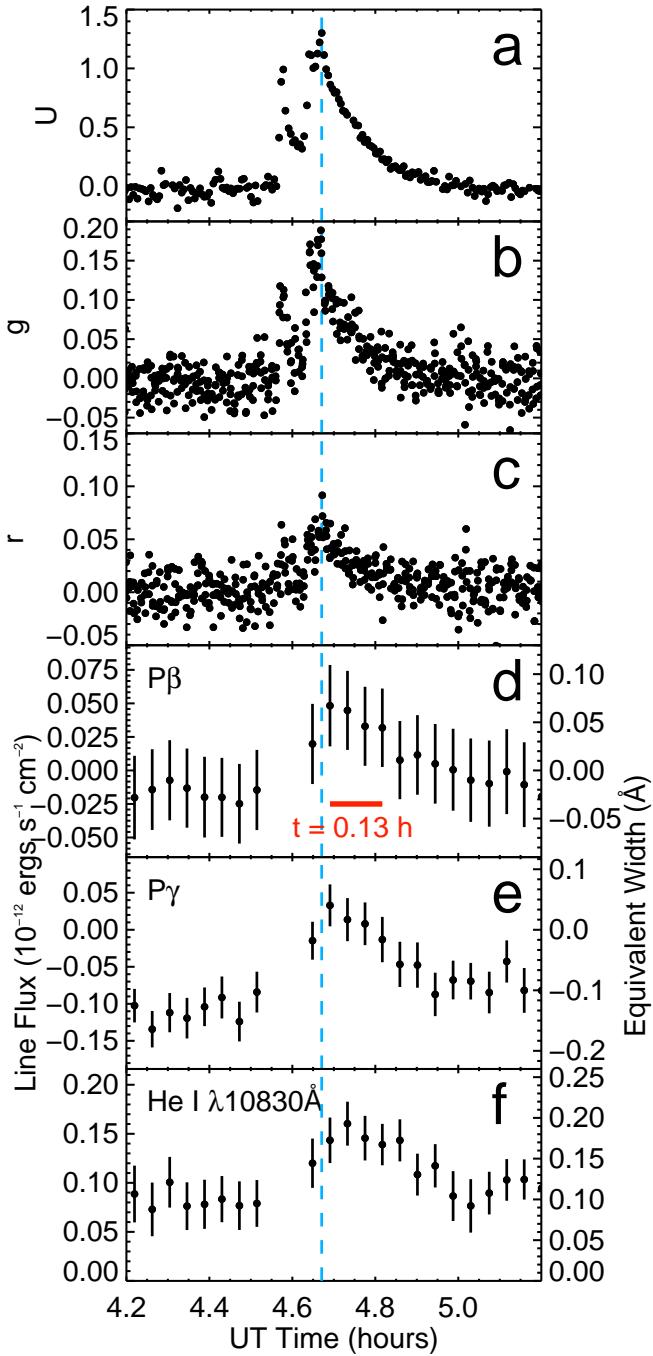} 
\caption{Data from the UT 2011 February 14 flare on YZ CMi. Panels (a), (b), and (c) show the change in $U$-, $g$-, and $r$-bands during the flare, and panels (d), (e), and (f) show the flux (left axis) and EW (right axis) of the infrared emission lines detected during the flare. Panel d shows the time used as the flare duration in Table~\ref{tab:targ} (red line). The time of the peak $u$-band flux is shown in each panel (blue dashed line). The infrared emission line fluxes are averaged over 16 exposures of 4 seconds each. The total exposure time for each point is 64 seconds; the average cadence is 2.5 minutes. The gap in the infrared line data is due to standard star observation.} \label{fig:yzfeb}
\end{figure}

\textbf{2011 February 14 flare on YZ CMi:}
On UT 2011 February 14, we observed a $\Delta U$ = 1.38 flare on YZ CMi with the NMSU 1-m, ARCSAT, and TripleSpec. The flare lasted for t = 0.5 hours and released a total $U$-band energy of 4.4 $\times$10$^{31}$ ergs. Figure~\ref{fig:yzfeb} shows the $U$, $g$, and $r$-band light curves ($i$-band was also observed but showed no change during the flare) and line fluxes from P$\beta$, P$\gamma$ and He I $\lambda$10803\AA. This is the lowest energy flare with any evidence of IR line emission, and the measured EW were small (0.05 to 0.2\AA), which provides a lower limit on the observability of IR line emission. With these small EW, it is difficult to compare the strengths of the emission lines; they are all the same strength within the uncertainties.

The $U$-band light curve shows a fast-rise exponential-decay shape with a precursor event 0.1 hours before the main peak. The IR emission does not show the precursor or the initial rise of the photometry. However, the co-added infrared measurements have an effective time-resolution of 2.5 minutes (due to the inclusion of time spent executing the nod pattern and readout), which is insufficient to resolve those features.

\subsection{How Often Does IR Line Emission Occur?}
In order to determine the expected rate, or duty cycle, of infrared line emission, we first defined detectable emission as approximately 1$\sigma$ above the mean quiescent level. The length of time with detectable emission is shown for each flare as the red horizontal line on the P$\beta$ light curve in Figures~\ref{fig:evoct}, \ref{fig:evnov} and \ref{fig:yzfeb} and given in Table~\ref{tab:targ}. The total time spent in emission for all three flares observed is 1.4 hours (out of 48.9 possible hours), which corresponds to an IR flare emission duty cycle of 2.8\%. Excluding VB 8, we calculate an IR flare emission duty cycle of 3.1\% (of 44.3 hours) for active mid-M dwarfs. 

We can also place a limit on the duty cycle using the flare frequency distributions from \citet{Hilton2011phd}, which give the number of $u$-band flares per unit time for each flare energy. The $u$ and $U$-band energies of the flares with accompanying infrared emission are all above $3\times10^{31}$ ergs, corresponding to a flare frequency $<$0.1 per hour. Multiplying this emission time per flare by the flares per hour gives a duty cycle of $<$4.6\%, in agreement with our independent estimate. A duty cycle of 2.8 - 4.6\% represents an upper limit on detectable emission at this S/N and resolution, as our criterion requires only a small detection in the brightest line. 

\section{Atmospheric Structure}
\label{sec:mod}
We used the static NLTE radiative transfer code RH \citep{Uitenbroek2001} to generate model spectra to compare with the emission lines observed in the UT 2009 October 27 flare on EV Lac. We calculated model spectra based on one-dimensional atmospheres, using a 20-level hydrogen atom, a 20-level calcium atom, and a 25-level helium atom. The multi-level atoms were required to generate the lines observed, while the simplification to a one-dimensional atmosphere allowed us to examine a larger range of chromospheric structures without the computationally intensive calculations required by a detailed treatment of flare physics \citep[e.g.,][]{Allred2006}.

\begin{figure}
\includegraphics[width=0.95\linewidth]{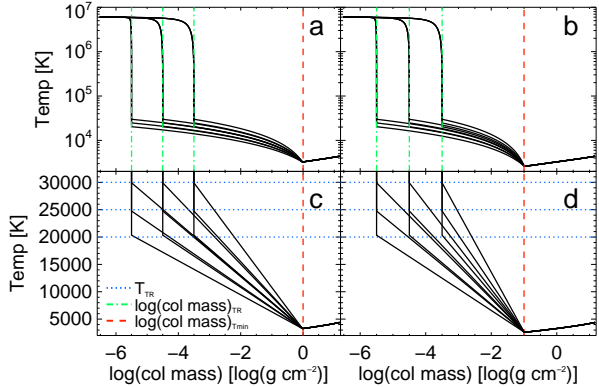} 
\caption{The atmospheric temperature structure as a function of log(col mass) for a subset of the atmospheres generated. Panels (a) and (b) show the entire atmosphere, while panels (c) and (d) focus on the chromosphere. The panels (a) and (c) show the models generated with a log(col mass)$_{Tmin} = 0$ and panels (b) and (d) show log(col mass)$_{Tmin} =-1$. Each panel shows 9 models, with a range of log(col mass)$_{TR} = -3.5$ to $-5.5$ (blue dot-dashed lines) and a range of $T_{TR} = $ $20\,000$K to $30\,000$K (dotted green lines). Line fluxes ratios for the spectra generated from these models are shown in Figure~\ref{fig:lfluxc}.} \label{fig:mod}
\end{figure}

For a starting atmosphere, we used a Nextgen photospheric model from a T = 3200K solar metallicity dwarf \citep{Hauschildt1999} and the corona of the pre-flare M dwarf atmosphere model of \citet{Allred2006}. Similar to \citet{Hawley1992}, \citet{Christian2003} and \citet{Fuhrmeister2010}, we used chromospheres with a linear temperature rise in log column mass (log(col mass)) to connect the photosphere and corona. The linear temperature rise is a simplification of the actual chromospheric structure during a flare, but it is useful for an initial investigation of the temperatures required to generate emission lines at each atmospheric height. To produce a suite of model atmospheres, we varied the column mass of the transition region (log(col mass)$_{TR}$), the column mass of the temperature minimum region (log(col mass)$_{Tmin}$), and the temperature of the chromosphere at the bottom of the transition region (T$_{TR}$). Figure~\ref{fig:mod} shows the temperature structure of a representative subset of the resulting atmospheres and illustrates the three quantities we varied.

Following \citet{Walkowicz2008b}, we adopted our initial ranges for T$_{TR}$, log(col mass)$_{TR}$, and log(col mass)$_{Tmin}$ from previous quiescent and flaring M dwarf chromosphere models \citep{Hawley1992, Mauas1994, Houdebine1997, Short1998b, Walkowicz2008a,Fuhrmeister2010}. Table~\ref{tab:param} shows the range of parameters adopted for each of these three quantities, which differ from previous parameter ranges only in T$_{TR}$. Previous model atmospheres have relatively constant T$_{TR}$ $\sim$ $10\,000$K, but our initial models with a range of T$_{TR}$ = $10\,000$K to $20\,000$K underproduced Paschen emission relative to Balmer emission, and showed a trend of increasing Paschen emission with greater T$_{TR}$. We increased the temperature of our hottest models to T$_{TR}$ = $30\,000$K in order to generate relatively more Paschen emission. Although T$_{TR}$ extends to hotter temperatures, it is consistent with results from the radiative-hydrodynamic simulations from \citet{Allred2006}, which show that material at the base of the transition region can be heated up to T = 10$^{6}$ K.

\begin{deluxetable}{lll} \tablewidth{0pt} \tabletypesize{\scriptsize}
\tablecaption{Model Atmosphere Parameters \label{tab:param}}
\tablehead{ \colhead{Parameter} & \colhead{Range} & \colhead{Best}  }
\startdata
log(col mass)$_{TR}$ & -5.5 to -3.5 & -5.5 to -4.5 \\
log(col mass)$_{Tmin}$ & -3 to 0 & 0 \\
T$_{TR}$ & $10\,000$K to $30\,000$K & $25\,000$K to $30\,000$K \\
T$_{min}$ & 2229K to 3264K & 3264K \\
\enddata
\end{deluxetable}

\begin{figure*}
\includegraphics[width=0.95\linewidth]{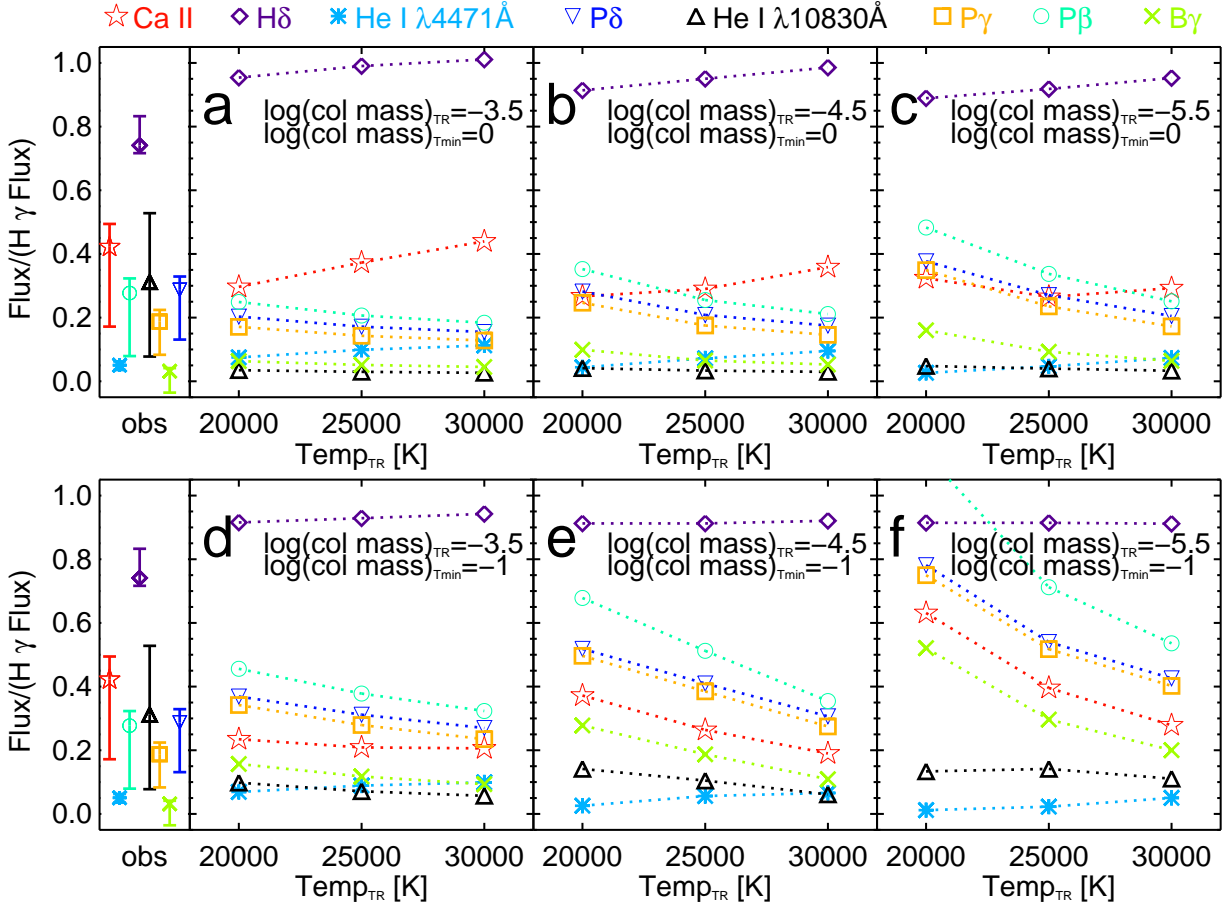} 
\caption{Line flux ratio to H$\gamma$ flux as a function of transition region temperature generated from the models shown in Figure~\ref{fig:mod}. Each line flux ratio is shown in a different color and symbol, which is detailed at the top. The two rows show the line flux ratios from models with different log(col mass)$_{Tmin}$. For comparison, the observed median and range of each line flux ratio is shown to the left of both rows. Each panel (a-f) shows models with a different log(col mass)$_{TR}$ and three values of T$_{TR}$. The dotted lines connecting the model line fluxes are shown only to clarify the positions of crowded points. The models that best represent the data are found in panels (a) and (b), with T$_{TR}$ = $25\,000$K and $30\,000$K.} \label{fig:lfluxc}
\end{figure*}

\begin{figure}
\includegraphics[width=\linewidth]{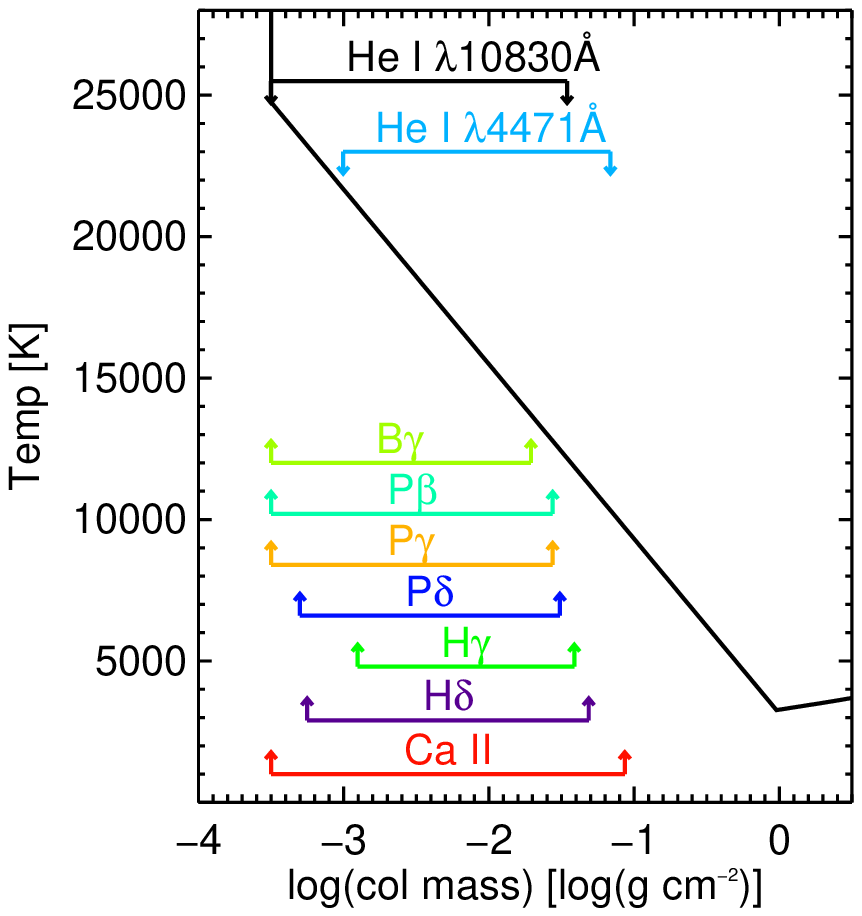}  
\caption{The atmospheric temperature structure as a function of log(col mass) for the best fit model (log(col mass)$_{Tmin} = 0$, log(col mass)$_{tr} = -3.5$, T$_{TR} = 25\,000$K). The range of formation depth is shown for each of the emission lines; the lines are grouped by type and offset in Temperature for clarity.} \label{fig:hgtfor}
\end{figure}

Comparing the strengths of the modeled lines to each other provides strong constraints on our suite of model atmospheres. The line flux ratios with respect to the H$\gamma$ line flux for the best models are shown compared to the median and range of observed line flux ratios in Figure~\ref{fig:lfluxc}. In general, a deeper T$_{min}$ (at log(col mass) = 0 or $-$1), a deeper transition region (at log(col mass) = $-$3.5 or $-$4.5), and a hotter chromosphere (with T$_{TR}$ = $25\,000$K or $30\,000$K) better reproduce the line flux ratios observed during the flare. The line formation regions (where the contribution function for each line is greater than 25\% of its peak value) for one model are shown in Figure~\ref{fig:hgtfor}.

Most of the lines are produced over regions that include the outer portion of the chromosphere, at log(col mass) = $-$3.5 and T = $25\,000$K. H$\gamma$, the strongest emission line we observed, is formed over the smallest portion of the chromosphere, with its highest temperature at T = $20\,000$K. He I $\lambda$4471\AA~and He I $\lambda$10830\AA~form over slightly different regions of the atmosphere, with He I $\lambda$10830\AA~tracing slightly higher temperatures.

During the flare observations, the ratio of H$\delta$ to H$\gamma$ is relatively constant. H$\delta$ is slightly overproduced in the models compared to observations, but is similar in each of the models. He I $\lambda$4471\AA~emission is weak (with a ratio to H$\gamma$ of $\sim$ 0.1 to 0.3)  in both the models and the observations.  The ratios of the Paschen series lines and Brackett $\gamma$ to each other are relatively constant throughout the flare, and those ratios are well produced in every atmospheric structure. The ratio of the Paschen lines and Br$\gamma$ to H$\gamma$, however, is matched only in the models with a log(col mass)$_{Tmin}$ = 0. Because these lines are sensitive to the hottest regions of the chromosphere, the advantage of the log(col mass)$_{Tmin}$ = 0 is likely an increased amount of material at temperatures near T = $20\,000$K due to a shallower slope in the chromosphere.

The ratio of Ca II K to H$\gamma$ and to the other hydrogen series lines is best produced in the models with the deepest T$_{min}$, a deep transition region (log(col mass)$_{TR}$ = -3.5 or -4.5), and a hot T$_{TR}$ = $25\,000$K or $30\,000$K. In all other models, Ca II K is underproduced relative to the Paschen series lines. As shown in Figure~\ref{fig:hgtfor}, the Ca II K emission in the best-fit model is formed over a larger range of log(col mass) than any other line. The production of Ca II K in a region that includes the upper chromosphere is unusual compared to previous results; typically, Ca II K emission during a flare is thought to last longer because it is a lower temperature line \citep{Houdebine2003,Crespo-Chacon2006}. Our cooler atmospheres, where Ca II K emission is formed only in lower temperature regions, do not produce enough Ca II K emission relative to Paschen series emission to match our observations.

He I $\lambda$10830\AA~is underproduced in nearly every model. During the flare, its observed ratio compared to H$\gamma$ increases from 0.1 to 0.5, while all our models show line flux ratios of 0.1 or less. This mismatch is apparently worse in the one of the other two flares observed; as discussed in Section~\ref{sec:flareall}, He I $\lambda$10830\AA~is stronger compared to P$\beta$ and P$\gamma$ (the two other lines observed) in the UT 2010 November 27 flare on EV Lac. Simply raising the T$_{TR}$ in our models produces too much Ca II K but no additional He I $\lambda$10830\AA. In the Sun and similar stars, He I $\lambda$10830\AA~emission is produced in the upper chromosphere during flares as a result of helium ionization via backwarming from coronal UV flux \citep{Mauas2005,Sanz-Forcada2008}. A similar process could be leading to the He I $\lambda$10830\AA~emission during M dwarf flares, but the details of backwarming from coronal emission are not yet fully implemented in the RH atmosphere code.

While one-dimensional atmosphere models can match the line flux ratios of most of the lines as a sequence of static snapshots, they cannot reproduce the time evolution of the flare. In our observations, Ca II K, the Paschen lines, Br$\gamma$, and He I $\lambda$10380\AA~all rise relative to H$\gamma$ during the decay phase of the flare. The best fitting models in our suite of atmospheres indicate an increase in Ca II K is always coupled with a decrease of the Paschen and Brackett lines. The time evolution of flares may involve different atmospheric components covering the surface of the star with changing filling factors \citep[e.g.,][]{Kowalski2010}. It is possible that a linear combination of two or three different one-dimensional atmospheres with changing filling factors would reproduce the time-evolution of this flare.

\section{Summary}
During nearly 50 hours of simultaneous photometric and spectroscopic observations on 4 active M dwarfs, we saw 16 total flares, 3 of them with accompanying infrared emission lines. The strongest flare ($\Delta u$ = 4.02) occurred on EV Lac on UT 2009 October 27. It showed emission from H$\gamma$, H$\delta$, He I $\lambda$4471\AA, Ca II K, P$\beta$, P$\gamma$, P$\delta$, Br$\gamma$, and He I $\lambda$10830\AA. A weaker flare ($\Delta u$ = 1.68) on EV Lac on UT 2010 November 27 showed only emission from P$\beta$, P$\gamma$, and He I $\lambda$10830\AA. Remarkably, the He I $\lambda$10830\AA~emission was twice as strong compared to P$\beta$ and P$\gamma$ as it was in the $\Delta u$ = 4.02 flare. The weakest flare with infrared emission ($\Delta U$ = 1.38) occurred on YZ CMi on UT 2011 February 14; P$\beta$, P$\gamma$, and He I $\lambda$10830\AA~were just above their detection limits. We estimate a duty cycle of 2.8\% to 4.6\% for observing the strongest infrared emission line (P$\beta$) during flares on active mid-M dwarfs. These observations confirm that flares are detectable in the infrared portion of M dwarf spectra, which is much brighter in quiescence than the bluer portions of M dwarf spectra which are typically used to detect flares. 

Using a hotter chromosphere than previous one-dimensional static flare models \citep[e.g.,][]{Christian2003,Fuhrmeister2010}, the ratios of Ca II K, He I $\lambda$4471\AA, H$\delta$, the Paschen lines, and Br$\gamma$ to H$\gamma$ can be relatively well reproduced. The generation of Ca II K in the hot, upper portion of the atmosphere is distinct from previous results, and is necessary to produce more Ca II K than Paschen series emission, which is observed during our strongest flare. This result confirms that infrared emission is a useful constraint on the atmospheric heating during M dwarf atmospheres.

The strength of emission from He I $\lambda$10830\AA~is not predicted from our one-dimensional model, but including a detailed treatment of backwarming from the corona \citep[e.g.,][]{Allred2006} may be warranted, based on solar results. Modeling He I $\lambda$10830\AA~is also complicated by its different emission strengths compared to P$\beta$ and P$\gamma$ in the two flares on EV Lac, but these differences show that He I $\lambda$10830\AA~has potential to constrain different backwarming scenarios during a variety of flares. The time-evolution of the largest flare is not reproduced by our one-dimensional models, but a combination of multiple models with different filling factors \citep[e.g.][]{Walkowicz2008b,Kowalski2010} or detailed radiative hydrodynamic modeling with non-thermal beam heating \citep[e.g.][]{Allred2006} may provide a better match to the flare emission.

\acknowledgments
We thank J. R. A. Davenport and H. Uitenbroek for helpful discussions and H. Uitenbroek for the use of his RH code. We also thank J. Holtzman for his assistance with the NSMU 1-m telescope and D. Monin for his help with the DAO 1.8-m. S. L. H., A. F. K., and E. J. H. acknowledge support from NSF grant AST 08-07205. J. P. W. acknowledges support from NSF Astronomy \& Astrophysics postdoctoral Fellowship AST 08-02230. B. M. T. acknowledges support from the Mary Gates Research Scholarship.

This publication is based on observations obtained with the Apache Point Observatory 3.5-meter telescope, which is owned and operated by the Astrophysical Research Consortium. This publication also makes use of data products from the Two Micron All Sky Survey, which is a joint project of the University of Massachusetts and the Infrared Processing and Analysis Center/California Institute of Technology, funded by the National Aeronautics and Space Administration and the National Science Foundation.

\bibliographystyle{apj}

\end{document}